\documentclass[10pt,conference]{IEEEtran}
\usepackage[ruled,vlined]{algorithm2e}
\usepackage{graphicx}
\usepackage{url}
\usepackage{paralist}
\usepackage{tabularx}
\usepackage{balance}
\usepackage{multirow}
\usepackage{multicol}
\usepackage{pbox}
\usepackage{mathtools}
\usepackage[TABBOTCAP]{subfigure}
\usepackage{algorithmic}
\usepackage{paralist}
\usepackage{tabularx}
\usepackage{balance}
\usepackage{multirow}
\usepackage{multicol}
\usepackage{amsmath}
\usepackage{setspace}
\usepackage{verbatim}
\usepackage{float}
\usepackage{xspace}
\usepackage[usenames]{color}
\usepackage{colortbl}
\usepackage{amssymb}
\usepackage{ifthen}
\usepackage{pifont}
\usepackage{tikz}

\include{macro}

\newcommand{\Fusionsp}{{\sc Fusion~}}

\ifCLASSINFOpdf
\else
\fi

\begin{document}

\title{On-Device Bug Reporting for Android Applications}

\author{\IEEEauthorblockN{Kevin Moran, Richard Bonett, Carlos Bernal-C\'ardenas, Brendan Otten, Daniel Park \& Denys Poshyvanyk}
\IEEEauthorblockA{College of William \& Mary\\
Department of Computer Science\\
Williamsburg, VA 23187-8795, USA\\
Email: \{kpmoran, rfbonett, cebernal, bwotten, dhpark01, denys\}@cs.wm.edu}}

\maketitle

\begin{abstract}
Bugs that surface in mobile applications can be difficult to reproduce and fix due to several confounding factors including the highly GUI-driven nature of mobile apps, varying contextual states, differing platform versions and device fragmentation.  It is clear that developers need support in the form of automated tools that allow for more precise reporting of application defects in order to facilitate more efficient and effective bug fixes.  In this paper, we present a tool aimed at supporting application testers and developers in the process of \textbf{O}n-\textbf{D}evice \textbf{B}ug \textbf{R}eporting.  Our tool, called \textbf{ODBR}, leverages the uiautomator framework and low-level event stream capture to offer support for recording and replaying a series of input gesture and sensor events that describe a bug in an Android application.
\end{abstract}

\section{Introduction}
\label{sec:intro}
% !TEX root = main.tex

	In order to aid in the often difficult process of reproducing and fixing bugs related to mobile apps, in this paper we present a tool that enables developers to effectively carry out the process of \textbf{O}n-\textbf{D}evice \textbf{B}ug \textbf{R}eporting. Our prototype \textbf{ODBR} app is capable of running on a standalone physical or virtual Android device and recording precise user touch interactions coupled to specific GUI-components as well as sensor data streams for a target app. Then, this information is sent to a Java web application that displays a detailed, actionable bug report including screenshots, and series of user events that can be replayed on a target device allowing developers and testers to debug the app.

\begin{figure*}[tb]
\centering
\vspace{-1.1cm}
\includegraphics[width=\linewidth]{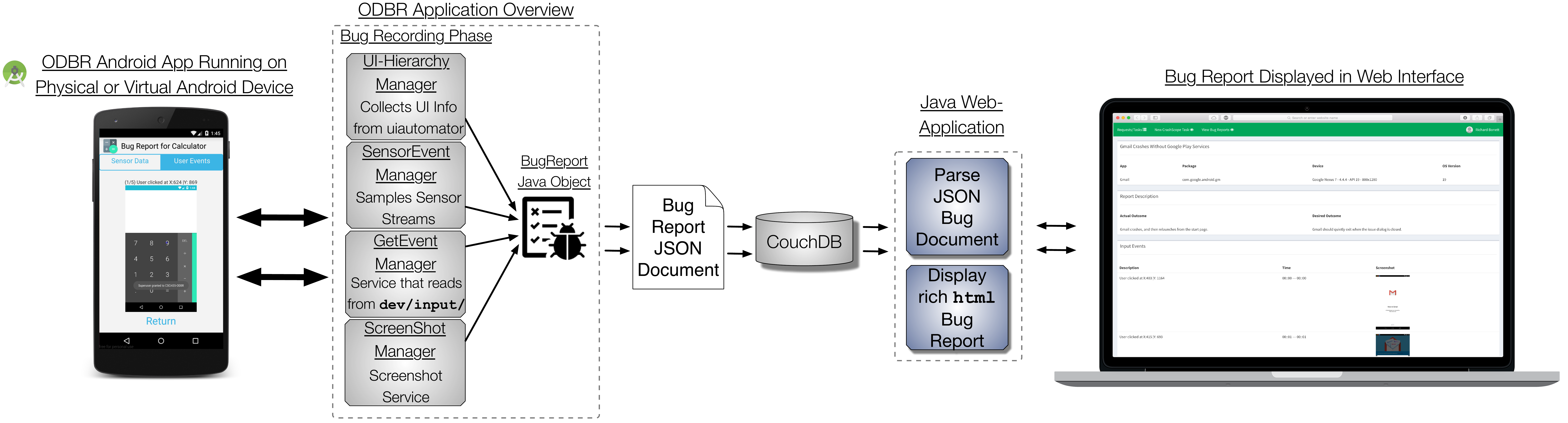}
\vspace{-0.4cm}
\caption{On-Device Bug Reporting Tool Architecture}
\vspace{-0.5cm}
\label{Design}
\end{figure*}

\section{Background and Related Work}
\label{sec:background}
% !TEX root = main.tex

	Given that effective bug and error reporting is a problem faced by nearly all Android developers, there are several existing commercial and research-oriented tools geared toward improving the reporting process.  Several commercial bug reporting and analytics platforms are available to support Android apps including Airbrake \cite{airbrake}, TestFairy \cite{test-fairy}, Appsee \cite{appsee}, and Bug Clipper\cite{bug-clipper}.  Typically, these solutions consist of an external library that a developer can include when writing their app. This library then enables the collection of information, such as device logs, performance information, screenshots and video recordings.  To the best of the authors' knowledge, none of these services is capable of collecting fine-grained user input and GUI information capable of being replayed on a target device as \textbf{ODBR} does. This precise collection methodology could capture critical information that a developer needs to successfully reproduce and fix an incoming bug report.

	Two recent automated input generation tools for mobile devices, CrashScope \cite{Moran:ICST16,Moran:ICSETD17} and Sapienz \cite{Mao:ISSTA16} are capable of producing detailed \textit{crash reports}. CrashScope uses a combination of static and dynamic analysis to perform targeted testing of application features according several touch, text, and contextual feature input generation strategies and is capable of producing an \texttt{html} crash report with images, and a replayable script.  Sapienz formulates the process of automated input generation as a multi-objective search problem, and is capable of producing videos and replayable scripts.  While these automated tools are able to provide relatively robust \textit{crash-reports} for apps they test, they are plagued by the oracle problem, in that while they can recognize and report when an app has crashed, they may miss user facing bugs that are more subtle.  Additionally, \Fusionsp \cite{Moran:FSE15, Moran:ICSE16} is an off-device bug reporting system for Android applications that leverages static and dynamic analysis performed before the reporting process to help guide users through reporting detailed reproduction steps for an application. \textbf{ODBR} is complimentary to \Fusionsp in that it provides another avenue for developers and testers to create detailed bug reports with minimal effort.  Finally, previous Record \& Replay approaches for Android apps have been proposed including RERAN \cite{Gomez:ICSE13}, VALERA \cite{Hu:OOPSLA15}, and MobiPlay \cite{Qin:ICSE16}.  While each of these solutions offers the ability to record user actions and replay them later, none of the tools offers the ability to record user actions on a standalone device without a connection to a host machine or server.  \textbf{ODBR} leverages the highly efficient and accurate event stream recording introduced by RERAN and VALERA and combines this with detailed GUI-hierarchy information collected via the \texttt{uiautomator} tool to enable the capture of precise information to aid in the debugging process.

\section{The ODBR Bug Reporting Tool}
\label{sec:approach}
% !TEX root = main.tex

 \textbf{The ODBR Workflow}: The overall architecture of our \textbf{ODBR} tool is illustrated in Figure \ref{Design}.  The entry point for a tester or developer using this tool is the \textit{ODBR Android Application}, which is available as an open source project accessible from our tool's webpage\footnote{\scriptsize \url{https://www.android-dev-tools.com/odbr}} along with further information about the project and demo videos. To use our tool, a reporter simply needs to install the \textbf{ODBR} app on their target virtual or physical device\footnote{\scriptsize Our tool's app currently requires a rooted target device, a requirement for accessing the \texttt{/dev/input} event stream}  select the app for which they wish to report a bug from the list of installed applications, hit the record button and perform the actions on the device that manifest a bug.  After the \textbf{ODBR} app (which is running in the background) detects that no touch-based inputs have been entered after a certain period of time, it asks the user if they have finished the report or if they wish to continue. Once the user has finished, they will have a chance to enter additional information about the bug, including a natural language description of the expected and actual behavior, and a title for the report.  Additionally, the user can view the screenshots and sensor data traces from the device and even replay the touch events to ensure they were properly captured. This replay feature relies on a custom written Java implementation of the \texttt{sendevent} utility\footnote{\scriptsize \url{https://goo.gl/EEUSfu}}. Once the report has been created, a \texttt{BugReport} is translated to a \texttt{json} document where it is sent over the web to a \texttt{couchDB} server instance. A Java web-application then reads the bug report information from the \texttt{json} file and converts the information into a fully expressive html report.  From the java web app, a developer can view the reproduction steps of the report, complete with screenshots and action descriptions, as well as download a replayable script that will reproduce the actions via \texttt{adb} or \texttt{sendevent} commands. 

 \textbf{ODBR App components}: The ODBR Application has 4 major components that aid in the collection of information during the bug reporting process.  These include: (i) the \textit{\textbf{GetEvent Manager}}, (ii) the \textit{\textbf{UI-Hierarchy Manager}}, (iii) the \textit{\textbf{Sensor Event Manager}} and (iv) the \textit{\textbf{Screenshot Manager}}.  The \textit{\textbf{GetEvent Manager}} is responsible for precisely and efficiently reading in the user event stream.  To accomplish this, during the reporting process the app creates threads that read from the underlying Linux input event streams at \texttt{/dev/input/...}  These input events provide highly detailed information about the user's interaction with the phone's physical devices, including (where applicable) the touch screen, power button, keyboard, etc. This information is identical to what is gathered using the Android \texttt{getevent} utility as used by RERAN \cite{Gomez:ICSE13}. Next, these low-level input event streams are parsed into higher level user interactions (e.g. \textit{swipe from (a,b) to (c,d)}). The low-level input events are retained to support precise analysis and replayability, while the higher level interactions are used to summarize the report in natural language. Whenever the \textit{\textbf{GetEvent Manager}} detects the user is taking a new action, it notifies the applicable managers to take a screenshot and dump of the UI-hierarchy, associating these with the new interaction. The \textit{\textbf{UI-Hierarchy Manager}} interfaces with the Android \texttt{uiautomator} framework to capture dumps of the Android view hierarchy in \texttt{ui-dump.xml} files for each new user action.  Because these dump files contain information about the screen location of each UI-component, we can use the event information obtained from the previous component to precisely infer the UI-component that the user interacted with at each step in the bug reporting process.  This component also extracts attributes of the various components on the screen including information such as the type (e.g., button, spinner) and whether the component is clickable. The \textit{\textbf{Sensor Event Manager}} is responsible for efficiently sampling the sensor input streams (e.g., accelerometer, GPS) during the bug reporting process.  This component accomplishes this by registering \texttt{SensorEventListener} instances for each sensor which sample the sensor values at appropriate rates.  Finally, the \textit{\textbf{Screenshot Manager}} is responsible for capturing an image of the screen for each new user interaction using the \texttt{screencap} utility included in Android.

\vspace{-0.1cm}
\balance
\bibliographystyle{abbrv}
\bibliography{ms}

\end{document}